\documentclass[twocolumn,secnumarabic,amssymb, nobibnotes, aps, prd]{revtex4-2}
\usepackage{mdframed}

\usepackage[export]{adjustbox}
\usepackage{graphicx}% Include figure files
\usepackage{dcolumn}% Align table columns on decimal point
\usepackage{bm}% bold math
\usepackage[normalem]{ulem}

\usepackage[title]{appendix}
\usepackage{bm}
\usepackage{listings,amsmath,amssymb,xcolor,graphicx,url}
\usepackage{mathtools}
\usepackage{natbib}
\setcitestyle{square, numbers}
\usepackage{subcaption} 
\usepackage{pgf}
\usepackage{svg}
\usepackage{import}
\usepackage[sc]{titlesec}  
\usepackage{caption} 
\usepackage{xcolor}

\usepackage[listings,skins]{tcolorbox}

\usepackage{notoccite}  

\definecolor{lgrey}{HTML}{808080}
\definecolor{dgrey}{HTML}{404040}

\definecolor{gnuplot-ls1}{HTML}{9400d3}
\definecolor{gnuplot-ls2}{HTML}{009e73}
\definecolor{web-green}{HTML}{00c000}

\allowdisplaybreaks

\usepackage{fancyhdr} 
\usepackage{lipsum}
\usepackage{setspace}
\graphicspath{./images/}  

\newcommand{\Rmin}{{R}_\text{min}}
\newcommand{\Rmax}{{R}_\text{max}}

\definecolor{dark-red}{rgb}{0.545 0 0}        
\definecolor{manc_purple}{RGB}{109,0,174}      
           
\newcommand{\yaw}{\chi_\textrm{yaw}}
\newcommand{\pitch}{\chi_\textrm{pitch}}
\newcommand{\roll}{\chi_\textrm{roll}}

\newcommand{\be}{\begin{equation}}
\newcommand{\ee}{\end{equation}} 
\newcommand{\bi}{\begin{itemize}}
\newcommand{\ei}{\end{itemize}}

\newcounter{subsubfigure}[subfigure]

\begin{document}

%revtex_shite \preprint{APS/??}

\title{From fluttering to drifting in inertialess sedimentation of achiral particles}

% Force line breaks with \\
\renewcommand{\andname}{\ignorespaces}

\author{Christian Vaquero-Stainer$^{1,2}$}
\author{Tymoteusz Miara$^{2}$}
\author{Anne Juel$^{2}$}
\author{Matthias Heil$^{3}$}
\author{Draga Pihler-Puzović$^{2}$}
\affiliation{%
  $^1$Nonlinear and Non-equilibrium Physics Unit, 
 Okinawa Institute of Science and Technology Graduate University, \\1919-1 Tancha, Onna, Okinawa 904-0495, Japan
}%
\affiliation{$^2$Department of Physics \& Astronomy and Manchester Centre for Nonlinear Dynamics, University of Manchester,  Oxford Road,  Manchester M13 9PL, UK}
\affiliation{$^3$Department of Mathematics and Manchester Centre for Nonlinear Dynamics, Oxford Road, Manchester M13 9PL, UK\\ }

\date{\today}% It is always \today, today,

\begin{abstract}
  There has been much recent interest in the chiral motion of achiral
  particles that sediment in a viscous fluid in a regime where
  inertial effects can be neglected. This occurs in a broad range of
  applications such as those involving biological objects like algae,
  ultra-thin graphene flakes, or colloidal suspensions. 
  It is known that particles
  with two planes of symmetry can be categorised as ``settlers'', ``drifters'' or
  ``flutterers'', where the latter sediment along chiral trajectories
  despite their achiral shapes. Here we analyse the sedimentation of
  circular disks bent into a U-shape (``flutterers'') and show how their behaviour
  changes when we break one of their symmetries by pinching the
  disks along their axis. The ``fluttering'' behaviour is found to be
  robust to such shape changes, with the trajectories now evolving
  towards helical paths. However, the behaviour changes when
  the degree of pinching becomes too strong, at which point the
  particles become ``drifters'' which sediment steadily without
  rotation. We establish criteria for the transition between the two
  types of behaviour and confirm our predictions
  in experiments. Finally, we discuss the implications of our
  observations for the dispersion of dilute suspensions made of
  such particles.
\end{abstract}

%revtex_shite\pacs{??}
%\pacs{46.70.De,47.15.km,62.30.+d,62.20mq}
\maketitle

%\tableofcontents

%--------------------------------------------------------------------------------------------------------
%	Introduction
%--------------------------------------------------------------------------------------------------------
\section{Introduction}
The motion of a rigid body through a viscous fluid under the influence of an
internal or external driving force is a canonical problem which is both
fundamental in nature and widely relevant to a range of applications.
The study of such problems dates back to seminal work of Stokes
\cite{Stokes51} who derived the drag on a sphere moving in the limit of vanishing
fluid inertia, a limit which is relevant for a host of biological and soft-matter
problems such as the size-separation
of biological molecules such as DNA~\cite{Weber13} and 2D
materials~\cite{Khan2012}, the provision of pathways to designing optimal
swimmers at the microscale~\cite{Keaveny13, Huseby24},
separating microplastics and other colloidal
suspensions~\cite{Doi2005, Marchetti2018}, and for
improving blood tests~\cite{Peltomaki13}.

At high Reynolds number, simple particles such as
rods, disks, spheres or ellipsoids moving in an unbounded
and otherwise quiescent fluid readily exhibit fluttering,
tumbling and chaotic modes, as observed for sedimenting
coins \cite{Field1997}, say. However, in the limit of zero Reynolds number,
such particles translate steadily without
reorientation~\cite{HappelAndBrenner,Stokes51, Brenner63}; and
more complicated dynamics only arise because of
the presence of bounding walls \cite{Mitchell2015},
background flows such as shear flows~\cite{Wang12,
  Thorp19}, or when multiple particles interact
hydrodynamically~\cite{Jung06, Salez15, Chajwa19}.

In an unbounded and quiescent Stokesian fluid, the sedimentation of a particle is
entirely determined by its shape~\cite{HappelAndBrenner}, and so 
non-trivial dynamics arise through translation-rotation
coupling, whereby translational motion
induces a torque which rotates the body, which, in turn, affect its translational
trajectory. It is this limit we are concerned with in the present study. The dynamics
are described by Stokes' theory which links the rate of change of the
particle's position and orientation to the forces and torques acting on it via the
so-called resistance matrix~\cite{HappelAndBrenner}.
Coupling between translation and rotation are encoded in the off-diagonal terms
in this matrix. 

The possible types of translation-rotation coupling depends on the
particle's symmetry \cite{HappelAndBrenner}; for example  it vanishes for
particles with three planes of symmetry, leading to simple translational motion.
For the case of chiral particles which do not have any planes of symmetry, such
as helices~\cite{Palusa18, Witten2020, Huseby24},
propellers~\cite{Tozzi11} and knots~\cite{Weber13}, 
their centres of mass tend to sediment along chiral trajectories,
whose handedness is determined by the particle shape. 

For achiral particles with fewer than three planes of
symmetry, the types of possible sedimentation trajectories can be determined by
analysis of their resistance matrices~\cite{Gonzalez2004,  Witten2020}. However,
there is no known general analytical approach to determine the resistance matrix
for an arbitrary particle shape, and so this must be determined
numerically or through experiments. The possible behaviour of an arbitrary
particle shape cannot, therefore, be determined \emph{a priori}.
Furthermore, even
when a particle exhibits non-trivial sedimentation dynamics, the
translation-rotation coupling can be so weak (as in the case of Kelvin’s
isotropic helicoid~\cite{Collins21}), that it would be difficult to
observe the behaviour in experiments. This is because the rotation
can occur over timescales during which the particles sediment
over much larger distances than can be accommodated in typical
lab experiments. 

We have recently performed experimental~\cite{Miara2024} and
numerical~\cite{Vaquero-Stainer24} studies of U-shaped disks
that were formed by isometrically warping a circular disk into
a cylindrical shape, resulting in a body with two planes of
symmetry. We showed that these achiral disks typically sediment in
a chiral manner, with the chirality determined
by their initial orientation rather than their shape. Such particles
are therefore examples of ``flutterers'', using the terminology
of~\cite{Joshi2025} whose recent theoretical study established the
possible types of inertialess sedimentation for particles
with two planes of symmetry. (In addition to ``flutterers'', this
class of particles includes ``drifters'' and ``settlers''). 

Here we combine numerical simulations and experiments 
to analyse how breaking one plane of the disk's symmetry affects this behaviour.  For this purpose we consider
the sedimentation of ``pinched'' U-shaped disks, obtained
by warping plane circular disks around a cone rather than a cylinder,
resulting in an achiral particle with a single plane
of symmetry; see Fig.~\ref{fig1}(a).
We show that variations in the degree of pinching
allow this particle to exhibit the entire range of possible behaviours,
from quasi-periodic spiralling (which is found to occur robustly over
a wide range of shapes) to sedimentation along straight
trajectories without any accompanying reorientation.

\begin{figure}
  \centering
   \vspace{-2mm} 
   \begin{subfigure}{0.35\linewidth}
     \hspace*{-5mm}
     \begin{tikzpicture}
       \draw (0, 0) node[inner sep=0] {
         \import{images/}{iso_conical_disk_matthias.tex}
       };
       \draw (-1.6,2.1) node {(a)};
     \end{tikzpicture}
    \vspace*{5mm}
    \caption*{}
  \end{subfigure}
  \vspace{-2mm}
  \begin{subfigure}{0.25\textwidth}
    \vspace{1mm}
   \begin{subfigure}{0.5\textwidth}
     %  \hspace*{0.3in}
     \begin{tikzpicture}
       \draw (0, 0) node[inner sep=0] {
         \import{images/}{euler_sequence_start.tex}
       };
       \draw (0.8,1) node {I};
       \draw (-0.5, 1) node {(b)};
     \end{tikzpicture}
    \end{subfigure}%
   \begin{subfigure}{0.5\textwidth}
     \hspace*{4mm}
     \begin{tikzpicture}
     \draw (0, 0) node[inner sep=0] {
         \import{images/}{euler_sequence_yaw.tex}
       };
     \draw (0.8,1) node {II};
     \end{tikzpicture}
    \end{subfigure} 
    \begin{subfigure}{0.5\textwidth}
      \hspace*{-2mm}
      \begin{tikzpicture}
     \draw (0, 0) node[inner sep=0] {
         \import{images/}{euler_sequence_pitch.tex}
       };
     \draw (0.8,1.35) node {III};
     \end{tikzpicture}
    \end{subfigure}%
    \begin{subfigure}{0.5\textwidth}
      \hspace*{2mm}
      \begin{tikzpicture}
     \draw (0, 0) node[inner sep=0] {
         \import{images/}{euler_sequence_roll.tex}
       };
     \draw (0.8,1.35) node {IV};
     \end{tikzpicture}
     \end{subfigure}
    \vspace{0.2mm}
     \caption*{}
  \end{subfigure}
  \caption{(a) ``Pinched'' disk constructed by the isometric
    deformation of a flat, circular disk onto the surface of a cone;
    (b) the intrinsic sequence of rotations II-IV, characterised by $\chi_{\rm
      yaw}$, $\chi_{\rm pitch}$ and $\chi_{\rm roll}$ (in that order)
    that reorientation of the disk from lab
    frame $\mathcal{F}_{\rm lab}$ to its general orientation in a
    body-fitted frame $\mathcal{F}_b$.
    \label{fig1}}
\end{figure}

%------------------------------------------------------------------------------
%	Methods
%------------------------------------------------------------------------------
\section{Mathematical Framework}
\label{sec:mathematical_framework}
We consider a thin disk of area $A=\pi {\cal R}^2$, nominal uniform thickness
$h \ll {\cal R}$ and density $\rho_d$, sedimenting
% in a finite container of size $L \gg {\cal R}$
under the action of gravity, $-g {\bf e}_3$, in a
Newtonian fluid of viscosity $\mu$ and density $\rho_f < \rho_d$ at a typical
velocity $\mathcal{U}=(\rho_d-\rho_f)g{\cal R}^2/\mu$, so that the
corresponding Reynolds number is small,
i.e. $Re=\rho_f\mathcal{U}{\cal R}/\mu\ll 1$, allowing us to neglect
inertial effects in the fluid. For modest density ratios, such that
$\rho_d/\rho_f \ Re \ll 1$, we can neglect the inertia of the disk
too, implying that the disk sediments quasi-steadily.

In this study we consider disks that are obtained by the isometric
deformation of planar circular disks of radius ${\cal R}$ onto the surface
of a cone, characterised by the two radii of curvature,
$({\cal  R}_{\rm min},{\cal R}_{\rm max}) ={\cal  R} (R_{\rm
  min},R_{\rm max})$. If $R_{\rm min} =
R_{\rm max}$ we obtain U-shaped disks with two planes of symmetry; otherwise
the disk is pinched, with its curvature varying with $x_2$, leaving
only a single plane of symmetry, as shown in Fig.~\ref{fig1}(a).
We refer to the most tightly curved point of a pinched disk as its ``nose''. 

We describe the disk's kinematics in terms of the
position vector $\textit{\textbf{r}}_M(t)$ to its
centre of mass (generally located outside the disk),
and characterise its orientation using an intrinsic
sequence of rotations defined by the Tait-Bryan angles
$\chi_{\rm yaw}(t)$, $\chi_{\rm pitch}(t)$ and $\chi_{\rm roll}(t)$,
with the rotations applied in that order, as shown in
Fig.~\ref{fig1}(b). Here $\chi_{\rm yaw} = \chi_{\rm pitch} =
\chi_{\rm roll} = 0 $ define the reference orientation in which the
disk's principal axes are aligned with the lab frame and the disk is
in an upright-U orientation.

%
% -----------------------------------------------------------------------------
\subsection{Resistance-matrix formalism}
% -----------------------------------------------------------------------------
%
Assuming that wall effects can be neglected, the disk's
quasi-steady motion is completely determined by the resistance
matrix $\mathbf{R}$ which links the disk's instantaneous
translational and rotational velocities, $d{\bf r}_{M}/dt$ and
$\pmb{\omega}$, to the net external force ${\bf F}$
and the torque ${\bf T}_M$ acting on its centre of mass,
\begin{equation}
      \left[
      \begin{array}{c}
        {\bf F} \\
        {\bf T}_{M}
      \end{array}
      \right]% ^{{\cal F}_b}
      = \mathbf{R} % ^{{\cal F}_b}
        %\underbrace{
        %\begin{pmatrix}
        %{\bf K} & {\bf C}_M \\
        %{\bf C}_M^T & {\bf \Omega}_M
        %\end{pmatrix}^{{\cal F}_b}
        %}_{\mbox{``Resistance matrix'' $\mathbf{R}^{{\cal F}_b}$}}
      \left[
      \begin{array}{c}
        d{\bf r}_{M}/dt \\
        \pmb{\omega}
      \end{array}
      \right]; % ^{{\cal F}_b};
    \label{formulation}
\end{equation}
see \cite{HappelAndBrenner}. If the vectors are decomposed into the
body-fitted coordinate system, ${\cal F}_b$, shown in
Fig.~\ref{fig1}(b), the entries of the
resistance matrix are constant and depend only on the shape of the
disk. For a freely sedimenting disk, the net force, ${\bf F}$,
is due to buoyancy in the direction of gravity which,
in the body-fitted frame ${\cal F}_b$, depends on the pitch and
roll angles, so ${\bf F}^{{\cal F}_b} =
{\bf F}^{{\cal F}_b}(\chi_{\rm pitch}(t),\chi_{\rm roll}(t))$. The
external torque is zero, ${\bf T}_{M}={\bf 0}$.

Equation (\ref{formulation}) represents a system of six
autonomous ODEs for the evolution of the three components of
${\bf r}_M(t)$ and the three Tait-Bryan angles $\chi_{\rm yaw}(t),
\chi_{\rm pitch}(t), \chi_{\rm roll}(t)$. In appendix \ref{app:governing_eqns}
we show that the ODEs for $\chi_{\rm pitch}$ and $\chi_{\rm
  roll}$ uncouple from the ODEs for $\chi_{\rm yaw}$ and ${\bf r}_M$
and have the form
\begin{equation}
\begin{split}
\frac{\rm{d} \chi_{\rm pitch}}{\rm{d}t} & = f_{\rm pitch}( \chi_{\rm pitch}, \chi_{\rm roll}), \\
\frac{\rm{d} \chi_{\rm roll} }{\rm{d}t} & = f_{\rm roll}( \chi_{\rm pitch}, \chi_{\rm roll}).
\end{split}\label{DynamicalSystem}
\end{equation}
Once $\chi_{\rm pitch}(t)$ and $\chi_{\rm roll}(t)$ have been determined from
(\ref{DynamicalSystem}), the other variables can be obtained
successively by integrating the remaining ODEs,
\begin{eqnarray}
\frac{\rm{d} \chi_{\rm yaw}}{\rm{d}t} & = & f_{\rm yaw}( \chi_{\rm
  pitch}, \chi_{\rm roll}),
\label{DynamicalSystem_chi_yaw}
 \\
\frac{{\rm d} {\bf r}_M}{{\rm d}t} & = & {\bf f}_M( \chi_{\rm yaw}, \chi_{\rm pitch}, \chi_{\rm roll}).
\label{DynamicalSystem_rm}
\end{eqnarray}
The functions on the right-hand-side of these ODEs are given in 
appendix \ref{app:governing_eqns}; they depend on the
coefficients of the resistance matrix which we computed using an
augmented finite-element-based solution of the 3D Stokes equations
(implemented in {\tt oomph-lib}~\cite{HeilHazelOomph2006}) to resolve the
singularities that develop along the edge of the disk;
see~\cite{Gupta57, Tannish96}. We refer to
\cite{VaqueroStainerThesis} and \cite{Vaquero-Stainer24} for
details of the implementation and validation.

The structure of equations (\ref{DynamicalSystem}),
(\ref{DynamicalSystem_chi_yaw}) and (\ref{DynamicalSystem_rm})
reflects the fact that in an unbounded
fluid, the rate at which the disk changes its inclination relative
to the direction of gravity (via changes to $\chi_{\rm pitch}$
and $\chi_{\rm roll}$) is not affected by changes to the yaw angle,
$\chi_{\rm yaw}$, or its position in the fluid, ${\bf r}_M$.

For later reference we note that
\begin{equation}
  \label{no_yaw_for_zero_roll}
  f_{\rm yaw}( \chi_{\rm  pitch}, \chi_{\rm roll}=0) = 
  f_{\rm yaw}( \chi_{\rm  pitch}, \chi_{\rm roll}=\pi) = 0,
\end{equation}
because a disk whose inclination is changed only by pitching does not
induce any rotation about the direction of gravity. We also have
\begin{equation}
  \label{yaw_and_roll_antisym}
  f_{\rm yaw}( \chi_{\rm  pitch}, \chi_{\rm roll}) = 
  - f_{\rm yaw}( \chi_{\rm  pitch}, -\chi_{\rm roll}).
  \end{equation}
Furthermore, the pinched disk's remaining plane of symmetry 
implies that
\begin{equation}
  \label{sym_in_f_pitch}
  f_{\rm pitch}(\chi_{\rm pitch}, \chi_{\rm roll}) =
  f_{\rm pitch}(\chi_{\rm pitch}, - \chi_{\rm roll})
\end{equation}
and
\begin{equation}
  \label{antisym_in_f_roll}
  f_{\rm roll}(\chi_{\rm pitch}, \chi_{\rm roll}) =
- f_{\rm roll}(\chi_{\rm pitch}, - \chi_{\rm roll}).
\end{equation}
%
% -----------------------------------------------------------------------------
\subsection{Two-dimensional phase plane}
% -----------------------------------------------------------------------------
%

Since $\chi_{\rm yaw}$ and ${\bf r}_M$ are enslaved to the evolution
of $\chi_{\rm pitch}$
and $\chi_{\rm roll}$, the system's evolution can be analysed
in a two-dimensional phase plane formed by these two angles.
\begin{figure}
  \hspace{-20mm}
  \begin{minipage}{0.5\textwidth}
    \hspace*{3mm}
    \import{images}{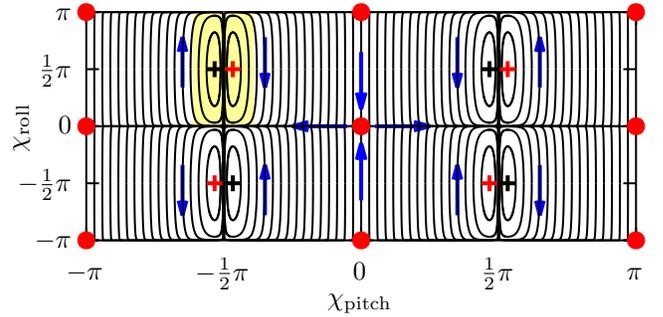}
  \end{minipage}
  \caption{Phase-space trajectories for a
    doubly-symmetric U-shaped disk with
    $\Rmin=\Rmax=1$. Filled red circles
    indicate the position of saddle-points, crosses indicate
    the location of centres, and arrows show the direction of travel
    along the paths. The colour of the crosses
    indicates the sense of rotation of the particles and the handedness of
    their helical trajectories in the lab frame:
    black [red] crosses indicate that particles rotate about the
    positive [negative] direction of gravity,
    ${\rm d}\chi_{\rm  yaw}/{\rm d}t < \ [ > ] \ 0$.
    \label{U-shaped_phase_plane}}
\end{figure}
Fig. \ref{U-shaped_phase_plane} provides a representative example:
it shows paths in the $(\chi_{\rm pitch},\chi_{\rm roll})$-phase plane
for a doubly-symmetric U-shaped disk with $R_{\rm min} = R_{\rm max}
=1$, the case studied in \cite{Vaquero-Stainer24}. In general, the
paths can be seen to be closed loops along which the disk
alternates between pitching and rolling-dominated
reorientations -- the behaviour exhibited by ``flutterers''.
We showed in \cite{Vaquero-Stainer24} that this is due to the fact that
the disk is stable to rolling but unstable to
pitching when it is in its upright-U orientation; conversely,
when the disk is in its upside-down-U orientation it is stable to
pitching but unstable to rolling. In the $(\chi_{\rm pitch},\chi_{\rm
  roll})$-phase plane the upright and upside-down-U orientations
(in which the disk sediments vertically without undergoing any
re-orientation, implying that for these isolated, unstable
orientations, the ``flutterers'' actually behave like
``settlers'') therefore appear as fixed points of
saddle type, represented by the filled red 
circles in Fig. \ref{U-shaped_phase_plane}. For example,
for $\chi_{\rm pitch}=\chi_{\rm roll}=0$, the $\chi_{\rm
  pitch}$ axis represents the saddle point's unstable directions
since a small perturbation to $\chi_{\rm pitch}$ continues to grow;
conversely the $\chi_{\rm
  roll}$ axis represents the stable direction since, following
a small perturbation to $\chi_{\rm roll}$,  the disk returns to its
upright orientation. (The lines $\chi_{\rm pitch} = \pm
\pi/2$ represent coordinate singularities, unavoidable for any
choice of Euler angles; along these, a change in
$\chi_{\rm roll}$ is exactly compensated for by an equal and
opposite change to the enslaved $\chi_{\rm yaw}$, so all points
on these lines represent the same orientation.)

The phase plane contains eight further fixed points,
$(\chi^*_{\rm pitch}, \chi^*_{\rm roll})$, which are
neutrally stable centres, identified by the crosses. When
released with this initial orientation, $\chi_{\rm pitch}(t), \chi_{\rm
  roll}(t)$ remain constant, implying that
the disk sediments without changing its inclination, but
rotating about the direction of gravity at
a constant rate ${\rm d}\chi_{\rm yaw}/{\rm d}t =
f_{\rm yaw}(\chi^*_{\rm pitch}, \chi^*_{\rm roll})$; see
(\ref{DynamicalSystem_chi_yaw}). This results in a helical centre-of-mass
trajectory with constant radius and pitch. The handedness of the
trajectories is indicated by the colour of the symbols
in Fig. \ref{U-shaped_phase_plane}: red [black] crosses represent
fixed points for which ${\rm d}\chi_{\rm yaw}/{\rm d}t$ is
positive [negative].

\begin{figure*}
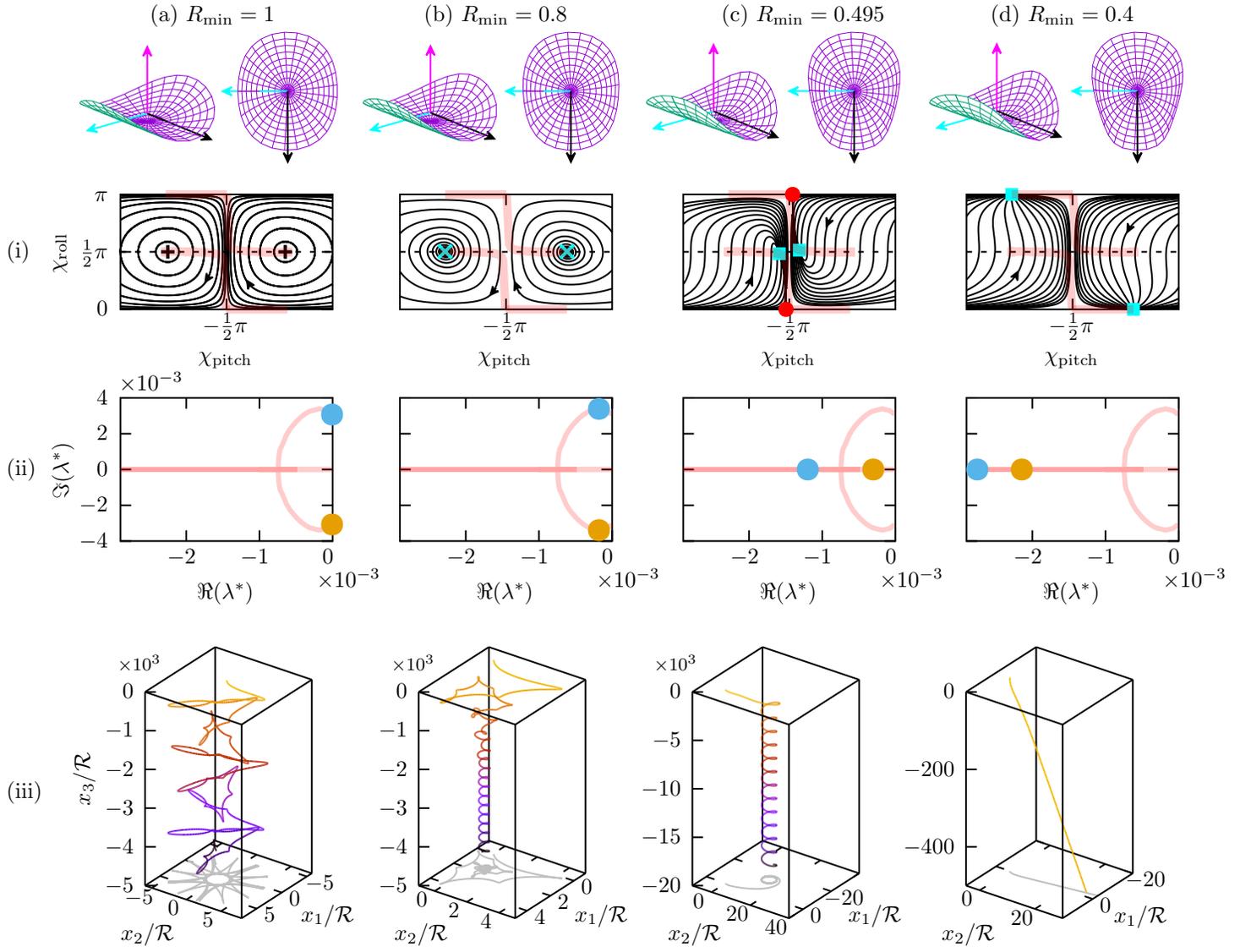

  \centering
  \begin{tabular}{cccc}
    \begin{tabular}{c}
      (a) $\Rmin=1$ \\[-1mm]
%      \hspace*{8mm}
      \import{images/}{conical_disk_Rmin_1.tex} \\[-4mm]
      \hspace*{-0.5in}
      \import{images/}{phase_portrait_Rmin_1_closeup_with_node_trajectory_paper.tex}      
      \\[-2mm]
      \hspace*{-0.5in}
      \import{images/}{eigenvalues_with_karaoke_points_R_min_1.tex}
    \end{tabular}
    &
    \begin{tabular}{c}
      \hspace*{-0.5in}
      (b) $\Rmin=0.8$ \\[-1mm]
      \hspace*{-0.5in}
      \import{images/}{conical_disk_Rmin_08.tex} \\[-4mm]
      \hspace*{-1.in}
      \import{images/}{phase_portrait_Rmin_08_closeup_with_node_trajectory_paper.tex}      
      \\[-2mm]
      \hspace*{-1.in}
        \import{images/}{eigenvalues_with_karaoke_points_R_min_08.tex}
    \end{tabular}
    &
    \begin{tabular}{c} 
       \hspace*{-0.2in}
       (c) $\Rmin=0.495$ \\[-1mm]
      \hspace*{-0.5in}
      \import{images/}{conical_disk_Rmin_0495.tex} \\[-4mm]
       \hspace*{-1in}
       \import{images/}{phase_portrait_Rmin_0495_closeup_with_node_trajectory_paper.tex}
       \\[-2mm]
       \hspace*{-1in}
       \import{images/}{eigenvalues_with_karaoke_points_R_min_0495.tex}
    \end{tabular} 
    &
    \begin{tabular}{c}
      \hspace*{-0.5in}
      (d) $\Rmin=0.4$ \\[-1mm] 
      \hspace*{-.5in}
      \import{images/}{conical_disk_Rmin_04.tex} \\[-4mm] 
      \hspace*{-1.in} 
      \import{images/}{phase_portrait_Rmin_04_closeup_with_node_trajectory_paper.tex}      
      \\[-2mm] 
      \hspace*{-1.in}
      \import{images/}{eigenvalues_with_karaoke_points_R_min_04.tex}
    \end{tabular} 
    \\[-4mm] 
    \multicolumn{4}{c}{
      \hspace*{-0.5in}
      \import{images/}{com_trajectories_vs_Rmin_wide_matthias.tex}
    }
  \end{tabular}
  \mbox{
    %\hspace{-16cm}
  \setlength{\unitlength}{1in}
  \begin{picture}(1,1)
    \put(-3.2, 5.65){(i)}
    \put(-3.2, 4.3){(ii)}
    \put(-3.2, 2.3){(iii)}
  \end{picture}
  }
 \vspace*{-1.2in}
 \caption{The behaviour of disks (with ${R}_{\rm max}=1$) with an
   increase in the degree of pinching in each successive column;
   (a) ${R}_{\rm min}=1$,
   (b) ${R}_{\rm min}=0.8$, ${R}_{\rm min}=0.495$,
   (c) ${R}_{\rm min}=0.4$. Row (i) shows the evolution of the
   disk's orientation in terms of pitch and roll angles in the part of
   the phase plane highlighted by the yellow region in
   Fig.~\ref{U-shaped_phase_plane}.  Arrows 
   illustrate the direction of the paths in the phase plane, and
   symbols indicate the locations of the fixed points: centres (black
   crosses), stable spiral nodes (cyan crosses), stable nodes (filled
   cyan squares) and saddle points (filled red circles). The
   translucent pink line shows the motion of the central fixed points
   through the phase plane as ${R}_{\rm min}$ is varied. Row (ii) show the eigenvalues $\lambda^*$ of the Jacobian matrix
   (\ref{jacobian}), evaluated at the central fixed points, with the
   translucent pink line illustrating their path through the complex plane
   as ${R}_{\rm min}$ is varied. Row (iii) shows the
   trajectories of the disk's centre of mass in the lab frame, with
   colours indicating the value of $x_3$; the projections of the
   trajectories into the $x_1-x_2$ plane is shown with gray lines.
 \label{big_pic}}
\end{figure*}

For all other initial orientations, $\chi_{\rm pitch}(t)$ and $\chi_{\rm
  roll}(t)$ change periodically as they traverse along the closed loops
in the phase plane. This leads to periodic changes in
${\rm d}\chi_{\rm yaw}/{\rm d}t$ which maintains its sign, resulting
in the disk rotating about the direction of gravity in the same 
direction as for the enclosed fixed point.  However, the
time it takes for $\chi_{\rm yaw}$ to increase by $2\pi$ is, in
general, not a rational multiple of the time it takes to traverse the
closed loop in the $(\chi_{\rm pitch}, \chi_{\rm roll})$-phase plane,
resulting in a quasi-periodic motion and complex spiral
trajectories in the lab-frame, as illustrated in Fig. \ref{big_pic}(a). 

\begin{figure}
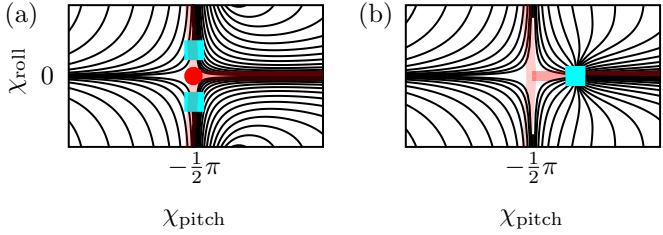

  \begin{subfigure}{0.25\textwidth}  
    \import{images/}{phase_portrait_Rmin_0491702_closeup_with_node_trajectory_paper.tex}
  \end{subfigure}%
  \begin{subfigure}{0.25\textwidth}
    \import{images/}{phase_portrait_Rmin_047_closeup_with_node_trajectory.tex}    
  \end{subfigure} %
%  \begin{subfigure}{0.25\textwidth}
%    \hspace*{-4mm}
%    \import{images/}{eigenvalues_with_karaoke_points_R_min_0491702.tex}    
%  \end{subfigure}%
%  \begin{subfigure}{0.25\textwidth}
%        \hspace*{-8mm}
  %    \import{images/}{eigenvalues_with_karaoke_points_R_min_047.tex}
%\end{subfigure} 
  \setlength{\unitlength}{1cm}
  \begin{picture}(1,1)
    \put(-3.8,3.9){(a)}
    \put( 0.9,3.9){(b)}
  \end{picture}
  \vspace{-10mm}
  \caption{\label{detail_of_pitchfork_in_phase_plane}Detail of the
    phase plane, showing the coalescence of the two stable nodes
    (filled cyan squares at $(\chi_{\rm pitch}^*, \pm \chi_{\rm
      roll}^*)$) with the saddle point (filled red
    circle at $\chi_{\rm roll} = 0$). Following the coalescence,
    only a single stable node remains at
    $(\chi_{\rm pitch}^*,\chi_{\rm roll}^* = 0)$.
    (a) $R_{\rm min} = 0.491702$  (b) $R_{\rm min} = 0.47$. These are
    between columns (c) and (d) in Fig. \ref{big_pic}.}
\end{figure}

\begin{figure}
\centering
    \hspace{-10mm}
    \import{images/}{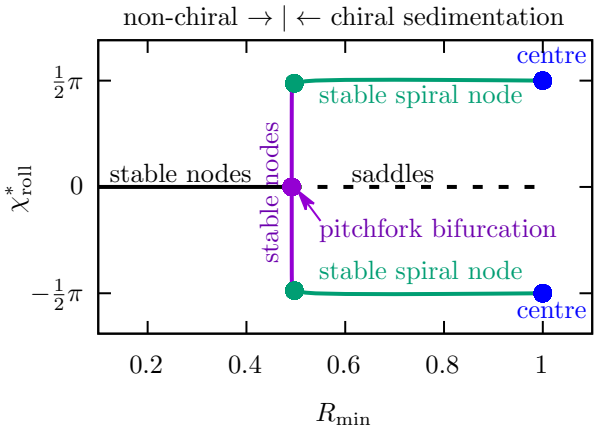}
  \caption{\label{bifurcation_for_chi_roll}Plot of
    the roll angle $\chi_{\rm roll}^*$ of the fixed point at which the
    disk (with $R_{\rm max}=1)$ sediments with a constant inclination as a
    function of $R_{\rm min}$. The character of
    the fixed point is indicated next to the curve. A nonzero value of
    $\chi_{\rm roll}^*$  implies that the sedimenting disk 
    rotates about the direction of gravity, resulting in a helical
    trajectory of its centre of mass.}
\end{figure}
%% %
%% b\egin{figure}
%%   \import{images/}{eigenvalue_real_part_vs_Rmin.tex}
%% \end{figure}
%% %

% -----------------------------------------------------------------------------
\section{Results}
% -----------------------------------------------------------------------------
%
\subsection{Numerical/theoretical results}
\label{subsec:numerical_results}
We will now investigate how this behaviour changes as we break the
disk's fore-aft symmetry by reducing $R_{\rm min}$ while keeping
$R_{\rm max} = 1$, thus pinching the disk along
its roll axis. For this purpose the figures in row (i) of
Fig. \ref{big_pic} show the sub-region of
the $(\chi_{\rm pitch},\chi_{\rm roll})$-phase plane
that we highlighted by the yellow shading
in Fig. \ref{U-shaped_phase_plane}. As before, black lines indicate the
paths of $\chi_{\rm pitch}(t)$ and $\chi_{\rm roll}(t)$ through the
phase plane, while symbols identify the position of the various
fixed points at which ${\rm d}\chi_{\rm pitch}/{\rm d}t=
{\rm d}\chi_{\rm roll}/{\rm d}t=0$. The translucent pink lines show
how the fixed points $(\chi^*_{\rm pitch},\chi^*_{\rm roll})$ (the
neutrally stable centres for the case of the
doubly-symmetric U-shaped disk) move through the phase plane
as  $R_{\rm min}$ is reduced from $R_{\rm min} = R_{\rm max} = 1$ (on the left),
to $R_{\rm min} = 0.4$ (on the right). Analytic expressions for these fixed points
are derived in appendix \ref{app:fixed_points}.

For $R_{\rm min}=1$ we see the behaviour already observed in
Fig. \ref{U-shaped_phase_plane}: the central fixed point in the phase
plane is a neutrally stable centre, and the eigenvalues $\lambda^*$ of the
Jacobian matrix
\begin{equation}
   {\bf J} =\left(
   \begin{array}{cc}
     \partial f_{\rm pitch}/\partial \chi_{\rm pitch} &
     \partial f_{\rm pitch}/\partial \chi_{\rm roll} \\
     \partial f_{\rm roll}/\partial \chi_{\rm pitch} &
     \partial f_{\rm roll}/\partial \chi_{\rm roll} \\
   \end{array}
   \right), \label{jacobian}
\end{equation}
evaluated at the fixed point, are purely imaginary as shown in row (ii) of Fig. \ref{big_pic}.

As we reduce $R_{\rm min}$, the fixed points
$(\chi_{\rm pitch}^*,\chi_{\rm roll}^*)$  move towards the line
$\chi_{\rm pitch}=-\pi/2$ and the associated eigenvalues become
complex conjugates with a negative real part, changing the character
of the fixed points from centres to stable spiral nodes. This
implies that, unlike
the case of the doubly-symmetric U-shaped disk, the disk's inclination now
evolves towards this fixed point, $(\chi_{\rm pitch},\chi_{\rm roll})
\to (\chi_{\rm pitch}^*,\chi_{\rm roll}^*)$, irrespective of its
initial orientation. As the inclination approaches the fixed point,
the disk's rate of rotation about the direction of gravity also
approaches a constant,
${\rm d}\chi_{\rm yaw}/{\rm d}t \to f_{\rm yaw}(\chi_{\rm pitch}^*,
\chi_{\rm roll}^*)$. Thus, following some initial transients, the
disk's centre of mass  ultimately sediments along a helical
trajectory, as shown in Fig. \ref{big_pic}(b.iii). Note that the character
of the initial transients is reminiscent of the behaviour of the
doubly-symmetric U-shaped disks: as the paths of $(\chi_{\rm
  pitch}(t), \chi_{\rm roll}(t))$ spiral towards their fixed point
$(\chi_{\rm pitch}^*,\chi_{\rm roll}^*)$ in the phase plane, the
disk's reorientation alternates between pitching and rolling-dominated
phases, with accompanying enslaved changes in the yaw angle. Hence,
during the transient phase, the trajectory of the disk's centre of
mass still follows a complex spiral path as it approaches 
its ultimate helical trajectory. We note that
the position of the saddle points in the phase plane
(indicated by filled red circles in Fig. \ref{U-shaped_phase_plane})
also changes with $\Rmin$, but for $\Rmin = 0.8$ they are still
outside the sub-region of the phase plane shown here. 
If the disk is released with an inclination that corresponds to
one of these saddle points it will in fact behave like a ``drifter'', i.e. it will
sediment downwards with a constant horizontal drift, but without any 
reorientation.  These trajectories are, of course, unstable, and so 
this behaviour will not be observable in experiments, justifying the
the decision of~\cite{Joshi2025} to base their classification into
``flutterers'', ``drifters'' and ``settlers'' on the particle shape 
rather than a particle's possible trajectories.

The plots in Fig. \ref{big_pic}(c) show that, as we
reduce $R_{\rm min}$ further, the two fixed points $(\chi_{\rm
  pitch}^*, \chi_{\rm roll}^*)$ move further 
towards the line $\chi_{\rm pitch}=-\pi/2$
and the complex conjugate eigenvalues coalesce on the real axis,
following which they initially move in opposite
directions: as $R_{\rm min}$ is reduced, the eigenvalue indicated by
the orange [blue] symbol in row (ii) of Fig. \ref{big_pic}
moves to the right [left] along the real axis.

When the eigenvalues become purely real, the stable spiral nodes
turn into stable nodes, identified by the filled cyan
squares in Fig. \ref{big_pic}(c.i). This means that the
approach of $(\chi_{\rm pitch}(t),
\chi_{\rm roll}(t))$ towards $(\chi_{\rm pitch}^*,
\chi_{\rm roll}^*)$ becomes monotonic rather than oscillatory,
resulting in the much simpler trajectory of the disk's centre of
mass during the initial transients: the disk no longer
alternates between pitching and rolling-dominated
reorientations and therefore approaches its final helical trajectory
without any complicated transient spiral motion; see the projection of the
centre-of-mass trajectory onto the $(x_1,x_2)$-plane
in Fig. \ref{big_pic}(c.iii). We note that the radius of the helical
trajectory increases with a reduction in $R_{\rm min}$.
Furthermore, two of the nine saddle points shown in
Fig.~\ref{U-shaped_phase_plane} have now moved so far along the
$\chi_{\rm roll}$ axis that they are now located close
to $\chi_{\rm roll} = -\pi/2$ (filled red circles).

As we reduce $R_{\rm min}$ further, the two fixed points $(\chi_{\rm
  pitch}^*,\chi_{\rm roll}^*)$ that have approached the line
$\chi_{\rm pitch}=-\pi/2$, move rapidly downwards [upwards]
towards $\chi_{\rm roll}=0 \  [\pi]$, where they approach the two saddle
points (shown in red) that have just moved into our sub-region of the phase
plane. Recall now that the symmetries of $f_{\rm pitch}$ and $f_{\rm  roll}$,
equations (\ref{sym_in_f_pitch}) and (\ref{antisym_in_f_roll}),
imply that each stable node at $(\chi_{\rm pitch}^*,
\chi_{\rm roll}^*)$ has a counterpart at $(\chi_{\rm
  pitch}^*,-\chi_{\rm roll}^*)$. The two close-ups in
Fig. \ref{detail_of_pitchfork_in_phase_plane} show that,
as $R_{\rm min}$ decreases, these two stable nodes
coalesce with the saddle point, resulting in the emergence of a
single stable node that then moves along the $\chi_{\rm roll}$ axis
as $R_{\rm min}$ is reduced yet further.

Equation (\ref{no_yaw_for_zero_roll}) shows
that $\chi_{\rm roll} = 0$ implies that
${\rm d}\chi_{\rm yaw}/{\rm d}t = 0$. Once the disk's
inclination has approached its fixed point $(\chi_{\rm pitch}^*,
\chi_{\rm roll}^*=0)$, the disk therefore ceases to rotate, and all three
Tait-Bryan angles remain constant.
Equation (\ref{DynamicalSystem_rm}) then implies that, following the
decay of the initial transients,  the disk's centre of mass sediments
along a straight trajectory, with constant velocity,
${\rm d}{\bf r}_M/{\rm d}t = \text{const.}$, and without any further re-orientation, as illustrated in 
Fig. \ref{big_pic}(d.iii).  The disk has thus turned from a ``flutterer''
into a ``drifter''. The transition from a helical to a straight
centre-of-mass trajectory  is achieved by the radius of the helix
becoming infinite as $\chi_{\rm roll}^*  \to 0$.

The coalescence of two stable nodes with a saddle point, illustrated
in Fig. \ref{detail_of_pitchfork_in_phase_plane},
indicates the occurrence of a (pitchfork) bifurcation, implying
that the Jacobian matrix (\ref{jacobian}) is singular and thus has
a zero eigenvalue. This corresponds to the eigenvalue indicated by the
orange symbol in the row (ii) of Fig. \ref{big_pic} reaching the
origin of the complex plane, following which it changes direction and
moves towards the left, like the other negative eigenvalue.

Fig. \ref{bifurcation_for_chi_roll} illustrates the
pitchfork character of the bifurcation using a plot
of $\chi_{\rm roll}^*$ as a function of $R_{\rm min}$. Reading this
plot from left to right, we start with a strongly pinched disk
for which the only fixed point is the stable node associated with $\chi_{\rm
  roll}^*=0$. As we increase $R_{\rm min}$ we reach the pitchfork
bifurcation (at $R_{\rm min} =  0.4977$; the filled magenta circle) beyond
which the disk has two stable inclinations, both with the same ``nose
down'' negative pitch angle, and equal and opposite values of
$\chi_{\rm roll}^*$. The magnitude of $\chi_{\rm roll}^*$ initially increases
very rapidly with an increase in $R_{\rm min}$ until it approaches the
limit $\chi_{\rm roll}^* \to \pm \pi/2$ which is attained for
the doubly-symmetric U-shaped disk characterised by
$R_{\rm min} = R_{\rm  max} = 1$. The point at which the character of the
fixed point changes from a stable node to a stable spiral node
is identified by the filled green circle. The transition occurs close
to the point where the rate of change of $\chi_{\rm roll}^*$ with
$R_{\rm min}$ changes rapidly, though there is no obvious causal
relation between these two events. 

\begin{figure*}
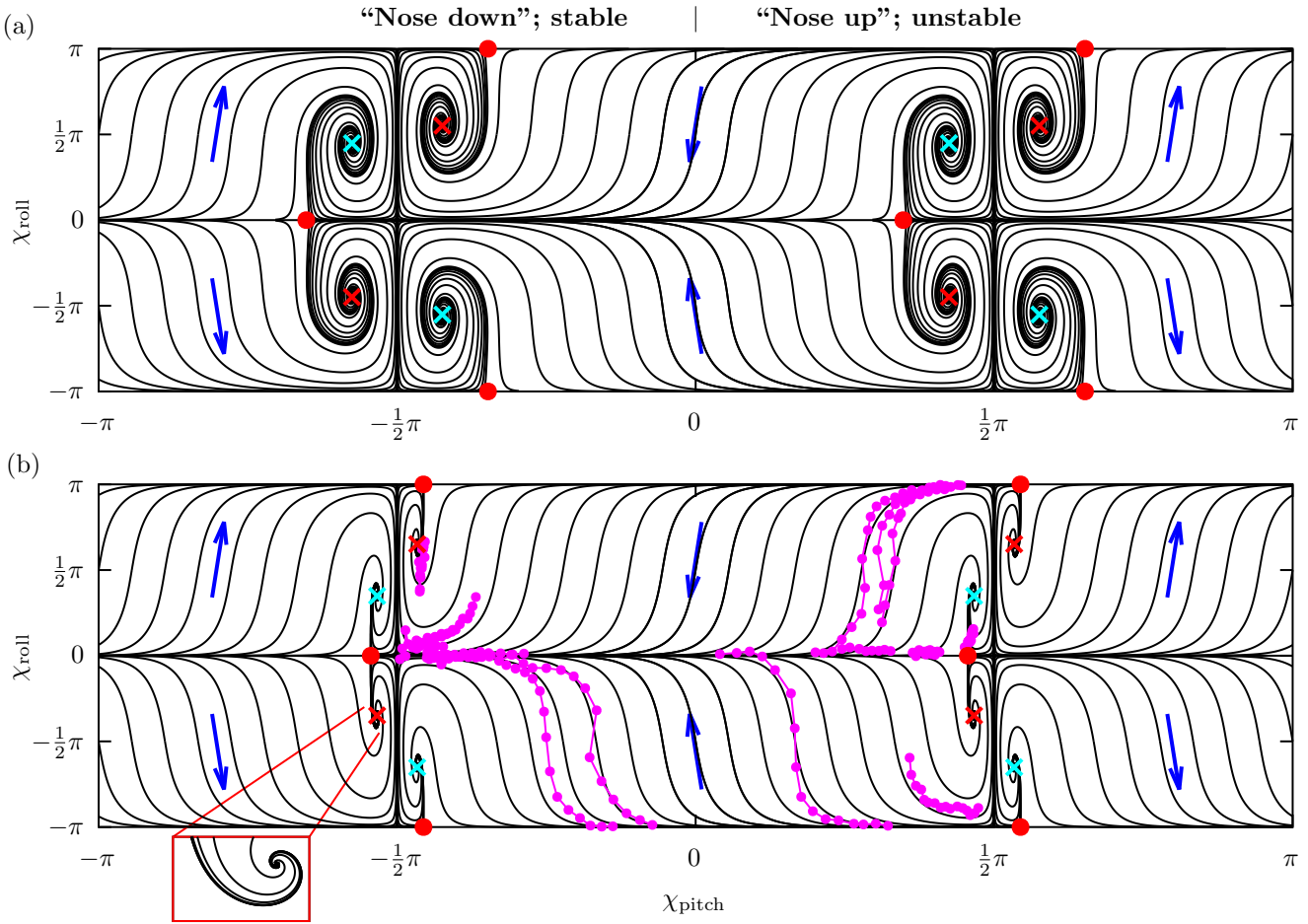
%[ht!] 
  \begin{subfigure}{\textwidth}
    \mbox{\normalsize \bf \color{black} \hspace{3mm}``Nose down'';
      stable \hspace{0.6cm} $|$ \hspace{6mm}``Nose up'';
      unstable}
    \vspace{-5mm}
    \\[2mm]
    \hspace*{-7in}(a)
    \vspace{-0.25in}
    \begin{center}
    \hspace{-10mm}
    \import{images/}{phase_portrait_tym_shape_theoretical.tex}
    \end{center}
  \end{subfigure}
  \\[-6mm]
  \hspace{-7in}(b)
  \\[-4mm]
  \begin{subfigure}{\textwidth}
    \begin{center}
      \hspace{-10mm}
      \import{images/}{phase_portrait_exp_comparison.tex}
    \end{center}
  \end{subfigure}
  \caption{\label{full_phase_with_experiments}Phase portrait for a
    disk with  $R_{\rm min} = 0.32, R_{\rm max} = 0.6$: Solid black
    lines show the paths in the phase plane, while the symbols
    identify fixed points where ${\rm d}\chi_{\rm pitch}/{\rm d}t =
    {\rm d}\chi_{\rm roll}/{\rm d}t = 0$. X-shaped symbols
    represent stable (left) and unstable (right) spiral nodes.
    The former correspond to disks sedimenting in a ``nose down''
    orientation while rotating about the direction of gravity  in
    a positive (cyan symbols) or negative
    (red symbols) sense. The corresponding points in
    the right half of the phase plane correspond to unstable fixed
    points in which the disk sediments in a ``nose up'' orientation.
    Filled red symbols represent saddle points.
    Trajectories and fixed points are computed via the framework in
    \S\ref{sec:mathematical_framework} using a resistance matrix computed via (a) FEM (as per
    \cite{VaqueroStainerThesis,Vaquero-Stainer24}); (b) fitting to experimental data (as per
    \cite{Miara2024}), with magenta symbols, connected by magenta lines, showing
    the raw experimental results; the red inset shows a closeup of one of the
    stable fixed points. Note that the geometry of the disk differed slightly from
    that shown in Fig. \ref{fig1}(a) to match the shape produced by
    the experimental fabrication method described in \cite{MiaraThesis}.}
\end{figure*}

So far we have restricted our attention to regions of the phase plane
in which the pinching of the disk turns the central fixed points from centres
into stable nodes. In Fig. \ref{full_phase_with_experiments}(a) we return
to the full phase plane and plot the paths $(\chi_{\rm pitch}(t),
\chi_{\rm roll}(t))$ for the case when $R_{\rm min} = 0.32, R_{\rm max}
= 0.6$, for which the central fixed points are spiral nodes. The plot
illustrates how the symmetries/anti-symmetries encoded in equations
(\ref{sym_in_f_pitch}) and (\ref{antisym_in_f_roll}) render the
four fixed points located in the region where $\chi_{\rm pitch} < 0$
(corresponding to the disk sedimenting with a ``nose
down'' inclination) stable. Conversely, the fixed points in the region
where the disk sediments with a ``nose up'' inclination, $\chi_{\rm
  pitch} > 0$, are unstable. The global phase portrait shows that the
four stable fixed points are globally attractive, so disks released with
any initial inclination ultimately evolve towards one of them.
However, apart from reflections (and the associated change of
handedness), the four fixed point actually
represent the same type of sedimentation:
the pairs $(\chi_{\rm pitch}^*,\pm\chi_{\rm roll}^*)$, correspond to 
disks that sediment with the same pitch angle but have rotated by equal
and opposite amounts about their roll axis. Equation
(\ref{yaw_and_roll_antisym}) shows that these disks rotate in opposite
directions about the direction of gravity, implying that
they sediment along helical centre-of-mass trajectories that
have the same shape (in terms of radius and pitch) but opposite
handedness. Furthermore, the orientations associated with the
two fixed points located ``diagonally across'' the points
$(\pm \pi/2,0)$ can be shown to have identical inclinations, with
their orientation only differing by a rotation about the yaw
axis by $\pi$.

Thus, following the decay of initial transients, sedimenting
pinched disks approach centre-of-mass trajectories of the
same shape  -- either helical or straight, depending on
the amount of pinching. Their initial orientation only
affects the handedness of the helical trajectories.

% -----------------------------------------------------------------------------
\subsection{Experiments}
% -----------------------------------------------------------------------------
%
We showed in \cite{Vaquero-Stainer24} that for doubly-symmetric
U-shaped disks (without any pinching), the prediction of
closed trajectories in the $(\chi_{\rm pitch},\chi_{\rm roll})$-phase
plane (as shown in Fig. \ref{U-shaped_phase_plane}) is
consistent with the experimental observations of \cite{Miara2024}.

Figs. \ref{full_phase_with_experiments}(a,b) show the corresponding
comparison for a pinched polyamide nylon disk of thickness $b
=237\,\mu$m, area $A = \pi {\cal R}^2 = 4.52\,$cm$^2$, curvatures
$R_\mathrm{min} = 0.32$ and $R_\mathrm{max} =0.6$, sedimenting
in a cuboidal tank of internal dimensions $90\times 40\times
40\,$cm$^3$. We note that the geometry of the disk used in the
experiments differed slightly from that shown in Fig. \ref{fig1}(a)
because of the way the disk was fabricated in the experiments;
see \cite{MiaraThesis}. Details of the
exact geometry (which was also used in the computations performed to
produce the data in Fig. \ref{full_phase_with_experiments}(a))
are provided in \cite{TymGithub}.

In the experiments we monitored the disk’s orientation while it sedimented
over a vertical distance of approximately $30{\cal R}$, staying at least
$10{\cal R}$ from the free surface and the side and bottom boundaries
of the tank.  The Reynolds number reached a maximum value of
$Re \sim 10^{-2}$, implying that inertial effects are likely to be negligible.  
Since the disk's reorientation takes place over very large vertical
distances (note the different vertical and horizontal scales in
Fig. \ref{big_pic}), each of the 15 experiments only provided a
relatively short segment of the paths in
the $(\chi_{\rm pitch},\chi_{\rm  roll})$-phase plane; these are
shown in magenta in Fig. \ref{full_phase_with_experiments}(b).
The black lines in that figure show the extrapolated behaviour,
obtained by fitting the entries of the resistance matrix to
the experimental data and then integrating equations
(\ref{DynamicalSystem}) from a range of initial
conditions; see ~\cite{Miara2024, MiaraThesis} for details.

Experiments and computations show that the symmetry breaking
induced by the pinching of the disk changes the stability of the
interior fixed points: instead of the orientation following neutrally-stable
periodic orbits as observed for doubly-symmetric U-shaped disks, the ``pinched''
disk approaches a constant inclination
in which it sediments along a helical trajectory in a nose
down orientation, irrespective of its initial
inclination. The extrapolated paths through the phase plane (black) agree well
with the short segments of raw experimental data (magenta) and allow the
identification of the position of the fixed points (saddle points, and
stable and unstable centres, all labeled as in
Fig. \ref{full_phase_with_experiments}(a)). In the experiments,
the fixed points are closer to the lines $\chi_{\rm pitch} = \pm
\pi/2$, but the topology of the paths in the phase plane, and thus the disk's 
overall behaviour, is consistent with that predicted by the computations. 

% -----------------------------------------------------------------------------
\section{Discussion}
% -----------------------------------------------------------------------------
%
We have investigated the inertialess sedimentation of a ``pinched'',
conically-deformed circular disk -- an achiral particle with a
single plane of symmetry --  utilising both experimental observations
and a theoretical analysis based on a resistance-matrix formalism.

It is known from our previous study \cite{Miara2024,Vaquero-Stainer24} that for
the special case where the disk is deformed onto a cylindrical surface
(resulting in a particle with two planes of symmetry), the disk
generally sediments along complex spiral trajectories whose shape and handedness
depend on the disk's initial orientation, the characteristic behaviour
exhibited by``flutterers''. For a particular
initial inclination, the trajectories of such disks are perfect helices.
Our current study showed that symmetry breaking by pinching makes
the helical trajectories globally attractive, so that they are
approached from all initial inclinations, with the disk sedimenting
in a ``nose down'' inclination while rotating about the direction of
gravity. (There are also isolated, unstable inclinations, associated
with the saddle points in the $(\chi_{\rm pitch},\chi_{\rm
  roll})$-phase plane, where the disks actually behave like
``drifters'' and thus sediment with a constant horizontal drift but without
re-orienting).  As the amount of pinching is increased, the radius of the
helical trajectories increases and ultimately becomes infinite,
resulting in straight trajectories along which the disk sediments
with constant velocity and without any reorientation, indicating
that the disk has become a ``drifter''.

 Our study shows that the sedimentation of achiral particles along
chiral trajectories is a remarkably robust phenomenon that persists
over a wide range of disk shapes and can readily be observed experimentally.
The fact that for sufficiently strong pinching the character of the
trajectories undergoes a qualitative change means that the conical
disks considered here span the range of behaviours from modes with
complex spirals trajectories to `drifting' modes where the trajectory of the
centre of mass is a straight lines inclined with respect to gravity.
The transition between the two regimes has
important consequences for the long-term behaviour of dilute
suspensions of such particles: suspensions made of particles
that move along spiral trajectories sediment with zero net
horizontal dispersion, whereas suspensions made of more
strongly ``pinched'' particles will spread out horizontally while
sedimenting without any rotation. This has the potential to be
exploited in applications that aim to control the collective
hydrodynamics of particles in such suspensions; see, e.g.,
~\cite{Witten2020}.

%{\bf *** do we need this data availability
%  shite? Extra hassle so avoid if we can. ***}
%\noindent\textbf{Data Availability:} The reported experimental and 
%numerical data are available from the Github repository: \\
%https://github.com/wobblygomboc/PinchedDiskData.

\begin{acknowledgements}
  C. Vaquero-Stainer and T. Miara were funded by EPSRC DTP studentships.
  A. Juel acknowledges funding by EPSRC (Grant EP/T008725/1).
\end{acknowledgements}
\bibliographystyle{plainnat}
\bibliography{paper}

\newpage

% -----------------------------------------------------------------------------
% -----------------------------------------------------------------------------
\appendix
\section{Derivation of the governing equations}
\label{app:governing_eqns}
In the absence of
inertia, the quasi-steady motion of a rigid particle in an unbounded
fluid is governed by a linear relation between the external force and
torque (about the centre of mass) acting on the particle,
and its resulting instantaneous rate of rotation and the velocity of
its centre of mass. In the following, we
non-dimensionalise all lengths on the disk's linear dimension ${\cal
    R}$; the velocities on the typical sedimentation velocity
  ${\cal U} = \Delta \rho g {\cal R}^2/\mu$, obtained by balancing
viscous and buoyancy forces; time on ${\cal R}^2/(\pi {\cal U} h)$;
  the rate of rotation on ${\cal R}/{\cal U}$; and the external force and
  torque on $\Delta \rho g {\cal R}^3$ and $\Delta \rho g {\cal R}^4$,
  respectively. Using tildes to distinguish non-dimensional variables
  from their dimensional counterparts, the relation between
  the external
  force $\widetilde{\bf F}$ and torque $\widetilde{\bf T}_M$ and the
  velocity of the particle's centre of mass, $\widetilde{\bf U}_M =
  {\rm d}\widetilde{\bf
    r}_M/{\rm d}\widetilde{t}$, and its
  rate of rotation $\widetilde{\pmb{\omega}}$ is then given by
  \begin{equation}
    \begin{bmatrix}
      \widetilde{\mathbf{F}} \\
      \widetilde{\mathbf{T}}_M 
    \end{bmatrix}= \widetilde{\mathbf{R}}
    \begin{bmatrix}
      \widetilde{\textbf{U}}_M \\
      \widetilde{\pmb{\omega}} 
    \end{bmatrix}.\label{formulation_appendix}
  \end{equation}
  If the vectors are decomposed into the body fitted coordinate
system ${\cal F}_b$, aligned with the particle's principal
axes (see Fig. \ref{fig1}), the entries in the resistance
matrix $\widetilde{\mathbf{R}}$ are
constant and depend only on the particle's shape. For particles that
have a reflectional symmetry about the $(x_2,x_3)$-plane in the
body-fitted coordinate system the matrix has the form
\begin{equation}
\widetilde{\mathbf{R}} = 
\begin{pmatrix}
           K_{11} & 0 & 0 & 0 & C_{12} & C_{13} \\
           0 & K_{22} & K_{23} & C_{21} & 0 & 0 \\
           0 & K_{23} & K_{33} & C_{31} & 0 & 0 \\
           0 & C_{21} & C_{31} &\Omega_{11} & 0 & 0 \\
           C_{12} & 0 & 0 & 0 & \Omega_{22} & \Omega_{23}\\
           C_{13} & 0 & 0 & 0 & \Omega_{23} & \Omega_{33}                
\end{pmatrix};
\end{equation}
see~\citep{HappelAndBrenner}. Here, $K_{ij}$ and $\Omega_{ij}$ are
the elements of the translation and rotation tensors, while the
coefficients $C_{ij}$ represent the coupling tensor which
characterises the tendency of the particle to rotate
when translating and vice versa.

In order to transform (\ref{formulation_appendix}) into quantities in the
lab-frame of reference, we recall that the disk's orientation
is described in terms of an intrinsic sequence of rotations by the
Tait-Bryan angles $\yaw, \pitch$ and $\roll$. 
Following the notation from Fig.~1(b) of the main text, this corresponds to 
\begin{equation}
   \scalebox{1.5}{${\cal F}_{\rm lab}$}  = \scalebox{1.5}{${\cal F}_0$} \stackrel{
     \scalebox{0.8}{$\yaw$ about $x_3^{{\cal F}_{\rm lab}}$}}{ \xrightarrow{\hspace*{2.05cm}} }
   \scalebox{1.5}{${\cal F}_{1}$}  \stackrel{
     \scalebox{0.8}{$\pitch$ about $x_2^{{\cal F}_{\rm 1}}$}}{
     \xrightarrow{\hspace*{2.05cm}} } \nonumber
\end{equation}
\begin{equation}
   \scalebox{1.5}{${\cal F}_{2}$}  \stackrel{
     \scalebox{0.8}{$\roll$ about $x_1^{{\cal F}_{2}}$}}{ \xrightarrow{\hspace*{2.05cm}} }
   \scalebox{1.5}{${\cal F}_{3}$} = \scalebox{1.5}{${\cal F}_{b}$}.
   \label{eqn:sequence_of_rotations}
\end{equation}
Hence, for an arbitrary vector  ${\bf a}$ with components $a_i^{{\cal F}_{j}}$ in
frame ${\cal F}_{j}$, its components
in the subsequent frame are given by
\begin{equation}
\bigg[
  a_{1}^{{\cal F}_{j+1}},
  a_{2}^{{\cal F}_{j+1}},
  a_{3}^{{\cal F}_{j+1}}
  \bigg]^T
= {\cal R}_{{\cal F}_{j}}^{{\cal F}_{j+1}} \ 
\bigg[
  a_{1}^{{\cal F}_{j}},
  a_{2}^{{\cal F}_{j}},
  a_{3}^{{\cal F}_{j}}
  \bigg]^T,
\end{equation}
where the matrices ${\cal R}_{{\cal F}_{j}}^{{\cal F}_{j+1}}$ are
standard orthogonal rotation matrices that depend only on the angle describing
the corresponding rotation. For example, 
\begin{equation}
{\cal R}_{{\cal F}_{0}}^{{\cal F}_{1}}
    =\left(
    \begin{array}{ccc}
      \cos(\yaw) & \sin(\yaw) & 0 \\
     -\sin(\yaw) & \cos(\yaw) & 0 \\
      0 & 0 & 1
      \end{array}
    \right) 
\end{equation}
and
\begin{equation}
    {\cal R}_{{\cal F}_{1}}^{{\cal F}_{0}} =
    \bigg({\cal R}_{{\cal F}_{0}}^{{\cal F}_{1}}\bigg)^{-1} =
    \bigg({\cal R}_{{\cal F}_{0}}^{{\cal F}_{1}}\bigg)^{T}.
\end{equation} 
Thus, translating the velocity of the disk's
centre of mass from ${\cal F}_{\rm lab}$ to
${\cal F}_{b}$ is achieved by
\begin{equation}
    \bigg[
  \widetilde{U}_{M,1}^{{\cal F}_{\rm b}},
  \widetilde{U}_{M,2}^{{\cal F}_{\rm b}},
  \widetilde{U}_{M,3}^{{\cal F}_{\rm b}}
  \bigg]^T
= {\cal R}_{{\cal F}_{\rm lab}}^{{\cal F}_{\rm b}} 
\bigg[
  \widetilde{U}_{M,1}^{{\cal F}_{\rm lab}},
  \widetilde{U}_{M,2}^{{\cal F}_{\rm lab}},
  \widetilde{U}_{M,3}^{{\cal F}_{\rm lab}}
  \bigg]^T= \nonumber
\end{equation}
\begin{equation}
 = {\cal R}_{{\cal F}_{\rm lab}}^{{\cal F}_{\rm b}}\bigg[
       \frac{{\rm d}\widetilde{r}_{M,1}^{{\cal F}_{\rm lab}}}{{\rm d}\widetilde{t}},
       \frac{{\rm d}\widetilde{r}_{M,2}^{{\cal F}_{\rm lab}}}{{\rm d}\widetilde{t}},
       \frac{{\rm d}\widetilde{r}_{M,3}^{{\cal F}_{\rm lab}}}{{\rm d}\widetilde{t}}
       \bigg]^T,
\end{equation}
where
\begin{equation}
   {\cal R}_{{\cal F}_{\rm lab}}^{{\cal F}_{b}} =
   {\cal R}_{{\cal F}_{2}}^{{\cal F}_{b}}\
   {\cal R}_{{\cal F}_{1}}^{{\cal F}_{2}}\
   {\cal R}_{{\cal F}_{\rm lab}}^{{\cal F}_{1}}.
\end{equation}   
Similarly, the rate of rotation is translated into the rate of change of
the Tait-Bryan angles via
\begin{equation}
\bigg[
  \widetilde{\omega}_{1}^{{\cal F}_{\rm b}},
  \widetilde{\omega}_{2}^{{\cal F}_{\rm b}},
  \widetilde{\omega}_{3}^{{\cal F}_{\rm b}}
  \bigg]^T
= 
{\cal R}_{{\cal F}_{2}}^{{\cal F}_{\rm b}} \ 
{\cal R}_{{\cal F}_{1}}^{{\cal F}_{2}} \
\bigg[ 0,0,\frac{{\rm d}\yaw}{{\rm d}\widetilde{t}} \bigg]^T
+ \nonumber
\end{equation}
\begin{equation}
+ {\cal R}_{{\cal F}_{2}}^{{\cal F}_{b}}\ 
\bigg[ \frac{{\rm d}\pitch}{{\rm d}\widetilde{t}}, 0, 0 \bigg]^T
+
\bigg[ 0,\frac{{\rm d}\roll}{{\rm d}\widetilde{t}}, 0 \bigg]^T 
 =  \nonumber
\end{equation}
\begin{equation}
  = {\cal S} \ \bigg[ 
  \frac{{\rm d}\pitch}{{\rm d}\widetilde{t}},
  \frac{{\rm d}\roll}{{\rm d}\widetilde{t}},
  \frac{{\rm d}\yaw}{{\rm d}\widetilde{t}}
\bigg]^T,
\end{equation}
where
\be
{\cal S} = 
\left(
\begin{array}{ccc}
  \cos(\roll) & 0 & -\cos(\pitch)\sin(\roll) \\
  0 & 1 & \sin(\pitch) \\
  \sin(\roll) & 0 &  \cos(\pitch)\cos(\roll) 
\end{array}
\right).
\ee
Thus, equation (\ref{formulation_appendix}) is transformed into
\be
\left(
\begin{array}{r}
     {\cal R}_{{\cal F}_{2}}^{{\cal F}_{b}}\
     {\cal R}_{{\cal F}_{1}}^{{\cal F}_{2}}\
     {\cal R}_{{\cal F}_{\rm lab}}^{{\cal F}_{1}}
     \bigg[
       \frac{{\rm d}\widetilde{r}_{M,1}^{{\cal F}_{\rm lab}}}{{\rm d}\widetilde{t}},
       \frac{{\rm d}\widetilde{r}_{M,2}^{{\cal F}_{\rm lab}}}{{\rm d}\widetilde{t}},
       \frac{{\rm d}\widetilde{r}_{M,3}^{{\cal F}_{\rm lab}}}{{\rm d}\widetilde{t}}
       \bigg]^T \\
     {\cal S} \
     \bigg[
       \frac{{\rm d}\pitch}{{\rm d}\widetilde{t}},
       \frac{{\rm d}\roll}{{\rm d}\widetilde{t}},
       \frac{{\rm d}\yaw}{{\rm d}\widetilde{t}}
       \bigg]^T 
\end{array}
\right) = \nonumber
\ee
\be
\label{big_system_in_lab_frame}
\mbox{\hspace{0cm}}
=\widetilde{\mathbf{R}}^{-1}\left(
\begin{array}{r}
     {\cal R}_{{\cal F}_{2}}^{{\cal F}_{b}}\
     {\cal R}_{{\cal F}_{1}}^{{\cal F}_{2}}\
     {\cal R}_{{\cal F}_{\rm lab}}^{{\cal F}_{1}}
     \bigg[
       0,
       0,
      -1
       \bigg]^T \\
     {\cal R}_{{\cal F}_{2}}^{{\cal F}_{b}}\
     {\cal R}_{{\cal F}_{1}}^{{\cal F}_{2}}\
     {\cal R}_{{\cal F}_{\rm lab}}^{{\cal F}_{1}}
     \bigg[
       0,
       0,
       0
       \bigg]^T
     \end{array}
\right).
\ee
Some straightforward if lengthy algebra then shows that the
evolution of $\chi_{\rm roll}$ and $\chi_{\rm pitch}$ is governed by
the system of two autonomous ODEs
\begin{equation}
\frac{{\rm d} \chi_{\rm pitch}}{{\rm d} \widetilde{t}} 
= \frac{1}{\mathbb{D}} (
          \cos(\chi_{\rm pitch}) \left(\mathbb{A} \cos^2(\chi_{\rm
            roll}) + \mathbb{B} \right) +  \nonumber
          \ee
          \be
          +\mathbb{C} \sin(\chi_{\rm
            pitch}) \cos(\chi_{\rm roll})),\label{DynamicalSystem1}
\end{equation}
\begin{equation}
\hspace{-2.5cm}\frac{{\rm d} \chi_{\rm roll}}{{\rm d} \widetilde{t}} = 
          \frac{1}{\mathbb{D}} \frac{\sin(\chi_{\rm roll})}{\cos(\chi_{\rm pitch})}
          (\mathbb{E} \cos^2(\chi_{\rm pitch}) +  \nonumber
          \ee
          \be
          \hspace{1cm}+\mathbb{A}
          \sin(\chi_{\rm pitch}) \cos(\chi_{\rm pitch}) \cos(\chi_{\rm
            roll}) + \mathbb{G} ).
          \label{DynamicalSystem2}
\end{equation}
With $\chi_{\rm pitch}(t)$ and $\chi_{\rm roll}(t)$ determined from
these equations, the evolution of the yaw angle then follows
from the solution of the ODE
\begin{equation}
  \frac{{\rm d} \chi_{\rm yaw}}{{\rm d}\widetilde{t}} = 
  - \frac{1}{\mathbb{D}} \frac{\sin(\chi_{\rm roll})}{\cos(\chi_{\rm pitch})}
  (\mathbb{H} \cos(\chi_{\rm roll}) \cos(\chi_{\rm pitch}) +
  \nonumber
  \ee
  \be
  + \mathbb{G} \sin(\chi_{\rm pitch})).
\end{equation}
Finally, the trajectory of the disk's centre of mass is determined by
%%%%%%%%%%%%%%%%%%%%%%%%%%%%%%%%
%
% DON'T EDIT THIS FILE! Regenerate by running
%
%    six_by_six_pinch.map in maple 
%
% and then run
%
%    turn_maple_output_into_latex.bash
%
%%%%%%%%%%%%%%%%%%%%%%%%%%%%%%%%
\begin{flushleft}\linespread{1.8}\selectfont \leftskip=3em\hspace{-3em}$\displaystyle \frac{{\rm d}\widetilde{r}_{M,1}^{{\cal F}_{\rm lab}}}{{\rm d}\widetilde{t}} = \big( (-2\mathbb{K}\mathbb{M}\sin(\chi_{\rm yaw})\cos(\chi_{\rm roll})\cos(\chi_{\rm pitch})^2+(-\mathbb{K}\mathbb{J}\sin(\chi_{\rm yaw})\sin(\chi_{\rm pitch})\cos(\chi_{\rm roll})^2+\mathbb{P}\cos(\chi_{\rm yaw})\sin(\chi_{\rm roll})\cos(\chi_{\rm roll})-(\mathbb{L}\mathbb{O}\sin(\chi_{\rm roll})^2-\mathbb{K}\mathbb{N})\sin(\chi_{\rm pitch})\sin(\chi_{\rm yaw}))\cos(\chi_{\rm pitch})+\mathbb{K}\mathbb{M}(\cos(\chi_{\rm roll})\sin(\chi_{\rm yaw})+\sin(\chi_{\rm roll})\sin(\chi_{\rm pitch})\cos(\chi_{\rm yaw}))) \big)  /(\mathbb{K} \mathbb{L}), $ \end{flushleft} 
\begin{flushleft}\linespread{1.8}\selectfont \leftskip=3em\hspace{-3em}$\displaystyle \frac{{\rm d}\widetilde{r}_{M,2}^{{\cal F}_{\rm lab}}}{{\rm d}\widetilde{t}} = \big( (2\mathbb{K}\mathbb{M}\cos(\chi_{\rm yaw})\cos(\chi_{\rm roll})\cos(\chi_{\rm pitch})^2+(\mathbb{K}\mathbb{J}\cos(\chi_{\rm yaw})\sin(\chi_{\rm pitch})\cos(\chi_{\rm roll})^2+\mathbb{P}\sin(\chi_{\rm yaw})\sin(\chi_{\rm roll})\cos(\chi_{\rm roll})+(\mathbb{L}\mathbb{O}\sin(\chi_{\rm roll})^2-\mathbb{K}\mathbb{N})\sin(\chi_{\rm pitch})\cos(\chi_{\rm yaw}))\cos(\chi_{\rm pitch})-\mathbb{K}\mathbb{M}(\cos(\chi_{\rm roll})\cos(\chi_{\rm yaw})-\sin(\chi_{\rm roll})\sin(\chi_{\rm pitch})\sin(\chi_{\rm yaw}))) \big)  /(\mathbb{K} \mathbb{L}), $ \end{flushleft} 
\begin{flushleft}\linespread{1.8}\selectfont \leftskip=3em\hspace{-3em}$\displaystyle \frac{{\rm d}\widetilde{r}_{M,3}^{{\cal F}_{\rm lab}}}{{\rm d}\widetilde{t}} = \big( -((\mathbb{J}\mathbb{K}\cos(\chi_{\rm roll})^2+\mathbb{L}\mathbb{O}\sin(\chi_{\rm roll})^2)\cos(\chi_{\rm pitch})^2-2\mathbb{K}\mathbb{M}\sin(\chi_{\rm pitch})\cos(\chi_{\rm roll})\cos(\chi_{\rm pitch})+\mathbb{K}\mathbb{N}\sin(\chi_{\rm pitch})^2) \big)  /(\mathbb{K} \mathbb{L}). $ \end{flushleft} 
   
The coefficients in these
ODEs depend on the entries in resistance matrix:
%%%%%%%%%%%%%%%%%%%%%%%%%%%%%%%%
%
% DON'T EDIT THIS FILE! Regenerate by running
%
%    six_by_six_pinch.map in maple 
%
% and then run
%
%    turn_maple_output_into_latex.bash
%
%%%%%%%%%%%%%%%%%%%%%%%%%%%%%%%%
\begin{flushleft}\linespread{1.8}\selectfont \leftskip=3em\hspace{-3em}$\displaystyle \mathbb{A} = ((C_{21}K_{23}-C_{31}K_{22})C_{13}^2+(-C_{21}^2K_{33}+2C_{21}C_{31}K_{23}-C_{31}^2K_{22}+\Omega_{11}(K_{22}K_{33}-K_{23}^2))C_{13}-\Omega_{33}K_{11}(C_{21}K_{23}-C_{31}K_{22}))\Omega_{22}+K_{11}(C_{21}K_{23}-C_{31}K_{22})\Omega_{23}^2-2((C_{21}K_{23}-C_{31}K_{22})C_{13}-\frac{1}{2}C_{21}^2K_{33}+C_{21}C_{31}K_{23}-\frac{1}{2}C_{31}^2K_{22}+\frac{1}{2}\Omega_{11}(K_{22}K_{33}-K_{23}^2))C_{12}\Omega_{23}+\Omega_{33}C_{12}^2(C_{21}K_{23}-C_{31}K_{22}),$ \end{flushleft}  
\begin{flushleft}\linespread{1.8}\selectfont \leftskip=3em\hspace{-3em}$\displaystyle \mathbb{B} = -(C_{12}\Omega_{23}-C_{13}\Omega_{22})((-K_{22}K_{33}+K_{23}^2)\Omega_{11}+C_{21}^2K_{33}-2C_{21}C_{31}K_{23}+C_{31}^2K_{22}), $ \end{flushleft}  
\begin{flushleft}\linespread{1.8}\selectfont \leftskip=3em\hspace{-3em}$\displaystyle \mathbb{C} = -(C_{21}K_{33}-C_{31}K_{23})((C_{13}^2-K_{11}\Omega_{33})\Omega_{22}+C_{12}^2\Omega_{33}-2C_{12}C_{13}\Omega_{23}+K_{11}\Omega_{23}^2), $ \end{flushleft}  
\begin{flushleft}\linespread{1.8}\selectfont \leftskip=3em\hspace{-3em}$\displaystyle \mathbb{D} = ((C_{13}^2-K_{11}\Omega_{33})\Omega_{22}+C_{12}^2\Omega_{33}-2C_{12}C_{13}\Omega_{23}+K_{11}\Omega_{23}^2)((-K_{22}K_{33}+K_{23}^2)\Omega_{11}+C_{21}^2K_{33}-2C_{21}C_{31}K_{23}+C_{31}^2K_{22}), $ \end{flushleft}  
\begin{flushleft}\linespread{1.8}\selectfont \leftskip=3em\hspace{-3em}$\displaystyle \mathbb{E} = (-C_{12}\Omega_{33}+C_{13}\Omega_{23})(-C_{21}^2K_{33}+2C_{21}C_{31}K_{23}-C_{31}^2K_{22}+\Omega_{11}(K_{22}K_{33}-K_{23}^2))-(C_{21}K_{33}-C_{31}K_{23})((-C_{13}^2+K_{11}\Omega_{33})\Omega_{22}-K_{11}\Omega_{23}^2+2C_{12}C_{13}\Omega_{23}-C_{12}^2\Omega_{33}), $ \end{flushleft}  
\begin{flushleft}\linespread{1.8}\selectfont \leftskip=3em\hspace{-3em}$\displaystyle \mathbb{G} = (C_{21}K_{33}-C_{31}K_{23})((-C_{13}^2+K_{11}\Omega_{33})\Omega_{22}-K_{11}\Omega_{23}^2+2C_{12}C_{13}\Omega_{23}-C_{12}^2\Omega_{33}), $ \end{flushleft}  
\begin{flushleft}\linespread{1.8}\selectfont \leftskip=3em\hspace{-3em}$\displaystyle \mathbb{J} = -C_{21}^2+K_{22}\Omega_{11}, $ \end{flushleft}  
\begin{flushleft}\linespread{1.8}\selectfont \leftskip=3em\hspace{-3em}$\displaystyle \mathbb{K} = (-C_{13}^2+K_{11}\Omega_{33})\Omega_{22}-K_{11}\Omega_{23}^2+2C_{12}C_{13}\Omega_{23}-C_{12}^2\Omega_{33}, $ \end{flushleft}  
\begin{flushleft}\linespread{1.8}\selectfont \leftskip=3em\hspace{-3em}$\displaystyle \mathbb{L} = -C_{21}^2K_{33}+2C_{21}C_{31}K_{23}-C_{31}^2K_{22}+\Omega_{11}(K_{22}K_{33}-K_{23}^2), $ \end{flushleft}  
\begin{flushleft}\linespread{1.8}\selectfont \leftskip=3em\hspace{-3em}$\displaystyle \mathbb{M} = -C_{21}C_{31}+K_{23}\Omega_{11}, $ \end{flushleft}  
\begin{flushleft}\linespread{1.8}\selectfont \leftskip=3em\hspace{-3em}$\displaystyle \mathbb{N} = -C_{31}^2+K_{33}\Omega_{11}, $ \end{flushleft}  
\begin{flushleft}\linespread{1.8}\selectfont \leftskip=3em\hspace{-3em}$\displaystyle \mathbb{O} = \Omega_{22}\Omega_{33}-\Omega_{23}^2, $ \end{flushleft}  
\begin{flushleft}\linespread{1.8}\selectfont \leftskip=3em\hspace{-3em}$\displaystyle \mathbb{P} = (((-K_{11}K_{22}+K_{22}K_{33}-K_{23}^2)\Omega_{33}+C_{13}^2K_{22})\Omega_{22}+(K_{11}K_{22}-K_{22}K_{33}+K_{23}^2)\Omega_{23}^2-2C_{12}C_{13}K_{22}\Omega_{23}+C_{12}^2K_{22}\Omega_{33})\Omega_{11}+((-C_{31}^2K_{22}+2C_{21}C_{31}K_{23}+C_{21}^2(K_{11}-K_{33}))\Omega_{33}-C_{13}^2C_{21}^2)\Omega_{22}+((K_{33}-K_{11})C_{21}^2-2C_{21}C_{31}K_{23}+C_{31}^2K_{22})\Omega_{23}^2+2C_{12}C_{13}C_{21}^2\Omega_{23}-C_{12}^2C_{21}^2\Omega_{33}. $ \end{flushleft}

\section{The fixed points}
\label{app:fixed_points}
The fixed points in the ($\chi_{\rm pitch},\chi_{\rm roll}$)-phase
plane are determined by the inclinations ($\chi_{\rm pitch}^*,\chi_{\rm
  roll}^*$) for which equations
(\ref{DynamicalSystem1}) and (\ref{DynamicalSystem2}) vanish
simultaneously. Given the product form of equation (\ref{DynamicalSystem2}),
the fixed points come in two families. For the first family,
equation (\ref{DynamicalSystem2}) vanishes because $\sin(\chi^*_{\rm
  roll})=0$. The corresponding solution of equation
(\ref{DynamicalSystem1}) depends on whether $\chi^*_{\rm
  roll}$ is an even or odd multiple of $\pi$, yielding the
two sets of fixed points
\begin{equation}
  (\chi^{*,I+}_{\rm pitch}, \chi^{*,I+}_{\rm roll})=
\left(\arctan\left(\frac{\mathbb{A}+\mathbb{B}}{\mathbb{C}}\right)+\pi
n, \ (2m+1)\pi\right),
\end{equation}
and
\begin{equation}
  (\chi^{*,I-}_{\rm pitch}, \chi^{*,I-}_{\rm roll})=
  \left(-\arctan\left(\frac{\mathbb{A}+\mathbb{B}}{\mathbb{C}}\right)+\pi
  n, \ 2m\pi \right)
\end{equation}
for any integers $n,m$. For $R_{\rm min} = R_{\rm max}$, these fixed
points are the saddle points represented by the red circles in
Fig. \ref{U-shaped_phase_plane}.

For the other family of solutions, equation (\ref{DynamicalSystem2}) vanishes
because its second factor is zero. This yields a further two
sets of fixed points
\begin{equation}
(\chi^{*,II\pm}_{\rm pitch}, \chi^{*,II\pm}_{\rm roll}) =
\left(\pm \arctan{(ab)}+n\pi, \ \arccos{(\pm b)} +
2m \pi \right), \label{centre_fixed_point}
\end{equation}
and their symmetric
counterparts at $(\chi^{*,II\pm}_{\rm pitch},-\chi^{*,II\pm}_{\rm roll})$,
again for any integers $n,m$. Here the coefficients $a$ and $b$ are
given by
\begin{equation}
  a =
  \frac{(\mathbb{G}+\mathbb{E})\mathbb{C}-
    \mathbb{B}\mathbb{A} - \sqrt{[(\mathbb{G}+\mathbb{E})\mathbb{C}-
        \mathbb{B}\mathbb{F}]^2+4\mathbb{A}\mathbb{B}\mathbb{G}(\mathbb{G}+
      \mathbb{E})}}{2\mathbb{B}\mathbb{G}}
\end{equation}
and
\begin{equation}
  b = \sqrt{-\frac{\mathbb{B}}{\mathbb{A}+a \mathbb{C}}}.
\end{equation}
For $R_{\rm min} = R_{\rm max}$, these fixed points are the centres
represented by the crosses in Fig. \ref{U-shaped_phase_plane}.

The pitchfork bifurcation occurs when the two fixed points described by equation
(\ref{centre_fixed_point}) approach their symmetric counterparts,
i.e. when $\chi^{*,II+}_{\rm roll} \to 0$ and $\chi^{*,II-}_{\rm roll} \to \pi$.
This occurs when $b \to 1$ which implies that
$a \to -(\mathbb{A}+\mathbb{B})/\mathbb{C}$.
Hence $(\chi^{*,II+}_{\rm pitch}, \pm \chi^{*,II+}_{\rm roll}) \to
(\chi^{*,I-}_{\rm pitch},\chi^{*,I-}_{\rm roll})$ and
 $(\chi^{*,II-}_{\rm pitch},\pm \chi^{*,II-}_{\rm roll}) \to
(\chi^{*,I+}_{\rm pitch},\chi^{*,I+}_{\rm roll})$, corresponding to the
coalescence between the two stable nodes and the saddle point described in
the main part of the paper.
  
\end{document}